\documentclass[10pt,onecolumn,amsmath,amssymb]{article}
\usepackage{amsfonts}
\usepackage{amsmath}
\usepackage{amssymb}
\usepackage{graphicx}
\usepackage{bm}
\usepackage{cite}
\usepackage{times}
\usepackage[font={small}]{caption}
\usepackage{mathrsfs}
\usepackage{relsize}
\usepackage{mathtools}

\newcommand{\onlinecite}[1]{\citen{#1}}

\makeatletter
\renewcommand\section{\@startsection{section}{1}{\z@}%
{-3.5ex \@plus -1ex \@minus -.2ex}%
{2.3ex \@plus.2ex}%
{\normalfont\bfseries}}
\renewcommand\subsection{\@startsection{subsection}{1}{\z@}%
{-3.5ex \@plus -1ex \@minus -.2ex}%
{2.3ex \@plus.2ex}%
{\normalfont\it\bfseries}}
\renewcommand\subsubsection{\@startsection{subsubsection}{1}{\z@}%
{-3.5ex \@plus -1ex \@minus -.2ex}%
{2.3ex \@plus.2ex}%
{\normalfont\it}}
\makeatother
\date{}
\renewcommand{\v}[1]{|{#1}\rangle}
\newcommand{\iv}[1]{\langle{#1}|}

\renewcommand{\O}{{\cal O}}
\newcommand{\lr}{{\leftrightarrow}}
\begin{document}

\author{\normalsize
Dmitry Solenov
\\\normalsize\it
Department of Physics, St. Louis University, St. Louis, Missouri  63103, USA
\\\normalsize\it
solenov@slu.edu
}

\title{\large\bf
Coherent modification of entanglement:
\\\large\bf
benefits due to extended Hilbert space
}

\maketitle\thispagestyle{empty}

\begin{abstract}
A quantum computing system is typically represented by a set of non-interacting (local) two-state systems---qubits. Many physical systems can naturally have more accessible states, both local and non-local. We show that the resulting non-local network of states connecting qubits can be efficiently addressed via continuous time quantum random walks, leading to substantial speed-up of multiqubit entanglement manipulations. We discuss a three-qubit Toffoli gate and a system of superconducting qubits as an illustration.

% Keywords:
\vspace*{10pt}
\vspace*{3pt}
%%%

\end{abstract}
\maketitle

\section{Introduction}

Quantum information is built around two-state quantum systems, qubits \cite{DiVincenzo,Chakrabarti,nielsenchuang}, which are manipulated by applying single- or few-qubit gates \cite{nielsenchuang,Kennedy,Strauch,Matsuo,Koch-Schoelkopf,Awschalom-SiC,Lu-Nori,Solenov-QDs,Carter,Solenov-NVs,Solenov-2QB}, and can be entangled as required by a quantum algorithm \cite{nielsenchuang,Shor-1,Grover-1,Shor-2,Cleve,Grover-2,Watrous}. In order to perform a sustainable quantum computation and benefit from quantum codes \cite{Shor-1,Shor-2,Grover-1,Cleve,Grover-2}, an error correction \cite{Steane,DiVincenzo-Shor,Kitaev-1} must take place to battle decoherence \cite{Leggett,vanKampen,Solenov-Privman-2,Solenov-STP}. Error correction, a quantum algorithm itself, rely on multiqubit entanglement and must be performed on a sufficiently coherent system to be successful \cite{Aharonov}. Therefore quantum operations involving multiple qubits must be sufficiently fast to enable continuous error correction. 

Although alternative concepts exist \cite{Mandel-Bloch,Browne-Rudolph,De-Pryadko}, in many physical implementations each qubit is a separate physical system \cite{Kennedy,Koch-Schoelkopf,Awschalom-SiC,Carter} isolated form other qubits. As the result, single qubit operations are easy to perform, while entangling operations are hard---the qubits must be set to interact temporarily. In this investigation we focus on systems where interactions are mediated via cavity modes (as, e.g., in superconducting transmon systems \cite{Koch-Schoelkopf}). In these systems multiqubit gates are performed via simpler one- and two-qubit gates \cite{nielsenchuang,Kitaev-1,Kitaev-2,Pham}. Thus, temporal complexity (speed of operation) of a typical entangling multiqubit gate \cite{nielsenchuang} grows rapidly with the number of qubits, $n$. Substantial effort has been made to reduce the gate operation time by optimizing two-qubit representations of multiqubit gates \cite{Shende,Markov,Maslov,Ionicioiu,Welch}. Although some speed-up (a constant factor) can be achieved this way \cite{Markov}, the complexity class---type of functional dependence on $n$---remains the same in most cases.

\begin{figure}\begin{center}
\includegraphics[width=0.8\textwidth]{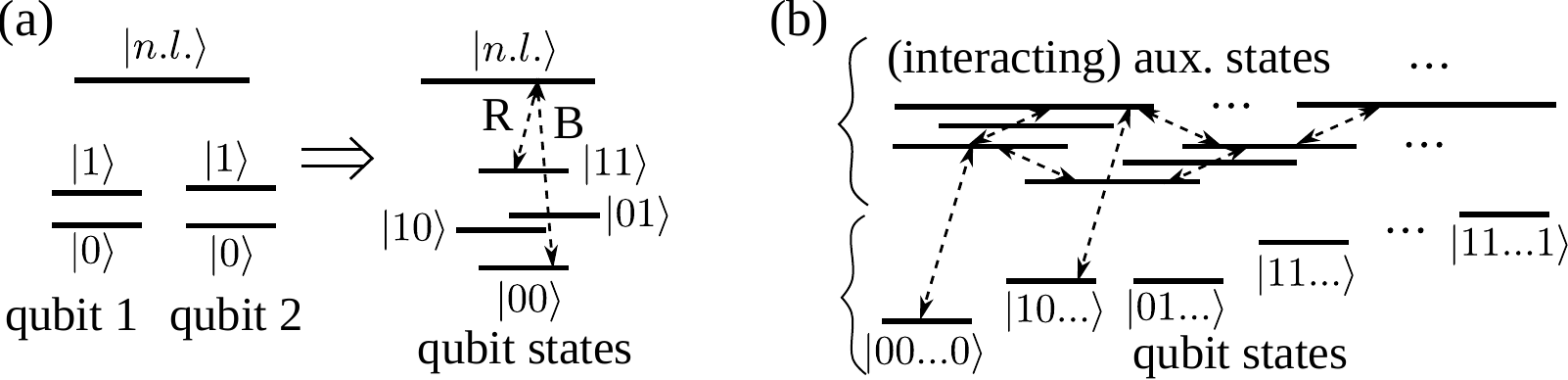}
\caption{\label{fig:MQB-scheme1}
Extended Hilbert space: (a) A schematic illustration of an entangling operation done via a non-local auxiliary state $\v{n.l.}$. (b) A schematic illustration of entangling operation on multiple qubits using multiple non-local auxiliary states to optimize (compress) the operation.
}\end{center}\end{figure}

We propose an approach to mutiqubit entangling operations that relies on accessible extended Hilbert space in which there is a set of always interacting (non-local) states involving multiple qubit systems. Unlike in the concept of qudits---multistate quantum units---we discuss excited states that can naturally remain interacting (non-local) when the qubit states are isolated from each other. Auxiliary excited states have been widely used to perform single- or two-qubit entangling gates in different qubit architectures \cite{Cirac,Blatt,Yale,Gammon1,Gammon2,Carter} via optical transitions using, e.g., $\Lambda$-system physics [see Fig.~\ref{fig:MQB-scheme1}(a)]. With this study we do not aim to suggest utility of non-local auxiliary states, which has been widely accepted \cite{Solenov-QDs,Solenov-NVs,Solenov-2QB,Cirac,Blatt,Gammon1,Gammon2}, rather we aim to demonstrate that the available structure of multiple auxiliary states [see Fig.~\ref{fig:MQB-scheme1}(b)] can be used to perform multi-qubit operations more effectively. Specifically, we show that multiqubit operations can be performed via quantum walks \cite{Farhi,Childs-Spielman,Shenvi,Kempe-REV} through such excited states: quantum data is still encoded in qubits, facilitating existing quantum codes, but the gate is performed by a quantum algorithm in an extended Hilbert space. We show that the combination of two factors: (i) {\it non-locality of higher-energy states} and (ii) representation of multiqubit gates via a {\it quantum walk}, can speed-up multiqubit operations potentially reducing their complexity class. Non-local states provide extra physical connections between qubits and quantum walks explore multiple quantum trajectories enabled by these connections at once. This approach can contribute to on-going research on gate compression schemes and can advance quantum algorithms, including error correction, reducing the fault-tolerance threshold. It can also be of interest in relation to other physical \cite{Brif,Zimbors,Dalibard} and biophysical \cite{Harel,Aspuru-Guzik-1,Harris-Kendon,Panitchayangkoon,Sarovar,Fassioli} problems.

Many physical implementations of quantum computing systems  have multiple well-defined quantum states that are routinely addressed in experiments \cite{Kennedy,Awschalom-SiC,Carter,Lucero-Martinis,Cirac,Blatt,Gammon1,Gammon2}. When such systems interact, these states form superpositions that are no longer localized in each individual system but can spread over a number of them [see Fig.~\ref{fig:MQB-scheme1}(b)]. The degree of non-locality of such superpositions is generally different at different energy levels.  This non-local network of states can be engaged by transferring quantum amplitudes from qubit states to higher energy states temporarily. As we will demonstrate, one of the most effective ways to accomplish this is to design continuous-time quantum walks \cite{Farhi,Childs-Spielman,Shenvi,Kempe-REV} through these networks of states. In what follows, we first introduce terminology. We, then, show how multiqubit gates can be performed via pulse sequences addressing always interacting auxiliary states. This approach, when optimized, can provide some gate compression but is limited because the structure of nonlocal states involving multiple qubits is not used. The pulse-sequence approach is used further as a benchmark to suggest increased efficiency of the proposed quantum-walk-based scheme.

\section{System and assumptions}
\label{sec:system}

In order to illustrate the concept we will focus on one of the simplest systems necessary: a collection of four-state quantum systems coupled to a single cavity mode of frequency $\omega_C$ [see Fig.~\ref{fig:MQB-scheme1}(b)]. Such configuration can be realized in a number of currently most advanced qubit architectures \cite{Lucero-Martinis,Cirac,Gammon1,Gammon2,Koch-Schoelkopf,Chow}. Generalization to larger number of states and other types of interaction is possible. We set each individual four-state qubit system (QS) to encode a qubit via the two lowest energy states $\v{i} = \v{0},\v{1}$, and use two additional states, $\v{e_i}=\v{e_0},\v{e_1}$, as auxiliary states. The Hamiltonian is
\begin{eqnarray}\label{eq:H}
H \!=\! H_{QSs} + \omega_C a^\dag a 
+ \!\sum_{i,n}\!\left(
g \v{i}^{(n)}\iv{e_i}^{(n)} a^\dag + h.c. 
\!\right)
+
V(t)
\end{eqnarray}
where
\begin{eqnarray}\label{eq:H_QSs}
H_{QSs} &=&  \!\!\sum_{i,n}\epsilon^{(n)}_i \v{i}^{(n)}\iv{i}^{(n)}
+ \sum_{i,n}\varepsilon^{(n)}_{i} \v{e_i}^{(n)}\iv{e_i}^{(n)}
\end{eqnarray}
Here constant energies $\epsilon_i$ and $\varepsilon_{i}$ correspond to states $\v{i}$ and $\v{e_i}$ respectively, $a^\dag$ ($a$) is the cavity mode creation (annihilation) operators, and the superscript $n$ enumerates QSs. External pulse control enters via $V(t)$. Other choices for cavity-qubit coupling, e.g., to $\v{1}\lr\v{e_0}$, can also be made \cite{Solenov-2QB}. This Hamiltonian (without $V$) can be diagonalized to
\begin{eqnarray}\label{eq:H_diag}
H = \sum_n E_n \v{\psi_n}\iv{\psi_n}
\end{eqnarray}
where states $\v{\psi_n}$ approach non-interacting states $\v{ij...}$, $\v{ie_j...}$, etc., when $g\to 0$, and index $n$ runs over the complete set of states as defined by Hamiltonian (\ref{eq:H}).

When all transitions in QSs are sufficiently detuned away from the resonance with $\omega_C$ ({\it off-resonance}) or have zero (or negligible) matrix elements, the cavity field can not mediate the interaction effectively. In this regime all transitions 
\begin{eqnarray}\label{eq:dW}
\v{i,jk...}\!\xleftrightarrow[]{A}\!\v{e_i,jk...},\quad \v{i,e_jk...}\!\xleftrightarrow[]{A'}\!\v{e_i,e_jk...},\quad ...
\end{eqnarray}
with the same $i$ are identical irrespective of the values of other indexes ($j,k,...$). This means that each QS is isolated and is not affected by the state of the rest of the system---only single-qubit manipulations are possible. 

When, on the other hand, the cavity mode is {\it in resonance} with at least one of the transitions in each QS and the corresponding matrix elements are non-negligible, the degeneracy can be lifted---a change in the state of one of the qubits can affect the entire system. As the result, entanglement can be modified, but single- (or few-) qubit operations are difficult to perform. We will label the superposition states $\v{\psi_n}$ at $g\neq 0$ the same way as the states from which they originate when $g$ is adiabatically increased from zero.

For the purpose of this investigation we will focus on the simplest case, when the cavity has non-negligible and identical matrix elements for transitions between states in each individual QS. This assumption simplifies the discussion because in this case it suffices to only discuss frequency resonance conditions and detunings that are independent of the details of qubit-cavity interactions in each individual physical system. On the other hand, this imposes the limit on how many qubits can be coupled this way until spectral crowding introduces unacceptable levels of cross-talk noise. This limit can be partially or completely lifted by using multiple cavities, each coupled to only few QSs, as will be briefly discussed at the end of this paper.

In the systems described by Hamiltonian of type (\ref{eq:H}) and with the assumptions stated above, a combination of both resonance and off-resonance limits---an {\it intermediate resonance regime} (IRR) \cite{Solenov-2QB}---is possible. In this regime, the cavity mode is somewhat away from resonance with every qubit system, and the decrease of qubit-cavity interactions (degree of non-locality) is different for different types of qubit-cavity states. Specifically, the cavity may be sufficiently far off resonance with QSs, such that transitions $A$ and $A'$ are indistinguishable for different $jk...$ and $k...$ respectively, while transitions in group $A$ remain distinguishable from those in group $A'$. We can quantify differences in the transition frequencies for transitions in each group by $\Delta\omega$, and differences in transition frequencies between transitions in group $A$ relative to that in group $A'$ by $\Delta\omega'$, and so on. We can also quantify the detuning of qubit transitions that belong to different qubits from each other by ${\frak O}$, i.e.,
\begin{eqnarray}\label{eq:Delta}
{\frak O}\sim |(\varepsilon^{(n)}_i-\epsilon^{(n)}_j)-(\varepsilon^{(n')}_i-\epsilon^{(n')}_j)|
\end{eqnarray}
for $n\neq n'$ and specific choice of $i,j$. When ${\frak O}/g\to\infty$, both $\Delta\omega$ and $\Delta\omega'$ vanish. The IRR is achieved at some finite values of ${\frak O}$ (and an appropriate choice of $\omega_C$ \cite{Solenov-2QB}) when
\begin{eqnarray}\label{eq:IRR}
\Delta\omega\ll\Delta\omega'.
\end{eqnarray}
While the IRR, as obtained in Ref.~\onlinecite{Solenov-2QB}, is specific to QS coupled via cavity modes, a similar structure with condition (\ref{eq:IRR}) can be present when interactions between QSs occur via other mechanisms.

\section{Pulse-controlled gates using a $\Lambda$-system}

In this section we give a brief overview of a standard $\Lambda$-system approach to two-qubit quantum gates to introduce concepts used in the other sections. This approach (and its variations) is used in different qubit architectures, e.g., in ion traps \cite{Cirac,Blatt} or in quantum dots \cite{Gammon1,Gammon2}, to manipulate entanglement. The key ingredient is a non-local auxiliary state that has non-zero matrix element for transitions involving at least one of the basis states in the two-qubit system, e.g., $\v{11}$, as in Fig.~\ref{fig:MQB-scheme1}(a). The state is non-local if it is formed as the result of interaction between qubits, either direct or mediated via a cavity field. The non-locality is manifested via the fact that transition $\v{11}\lr\v{n.l.}$ is different and distinguishable from $\v{10}\lr\v{n.l.}$ or $\v{01}\lr\v{n.l.}$. In order to perform the simplest entangling operation, a CZ (Control-Z) gate, it suffices to send a $2\pi$ pulse in resonance with $\v{11}\lr\v{n.l.}$ transition. Resonance pulses perform a rotation according to
\begin{eqnarray}\label{eq:Lam:2pi}
\v{\psi} = \cos(\theta/2)\v{11} - ie^{i\phi}\sin(\theta/2)\v{n.l.}
\end{eqnarray}
where $\theta$ is the integral of the pulse amplitude. When $\theta=2\pi$ the amplitude is returned back to the original state in the computational basis with a phase factor of $\pi$. As the result, the initial two-quit state is altered as
\begin{eqnarray}\label{eq:Lam:psi}
A\v{00}+B\v{01}+C\v{10}+D\v{11}
\to
A\v{00}+B\v{01}+C\v{10}-D\v{11}
\end{eqnarray}
which is equivalent to application of the CZ gate. Several pulse fields using both legs of the $\Lambda$-system can be used to create more complex entangling two-qubit rotations \cite{Gammon1,Gammon2}. Despite the simplicity of this scheme, the existence of suitable ``bus'' state $\v{n.l.}$ in systems with multiple qubits is hardware-dependent and may be difficult to have permanently in many architectures.

\section{Pulse-controlled gates in IRR}

\begin{figure}\begin{center}
\includegraphics[width=0.7\columnwidth]{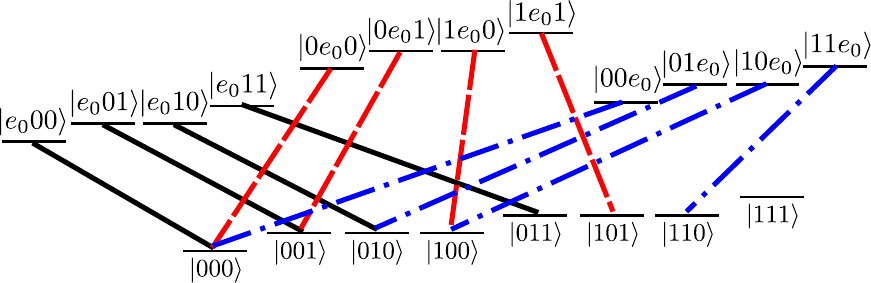}
\caption{\label{fig:3QB-MQB-A}
All $\v{0}\lr\v{e_0}$ transitions in a three qubit system (schematically). Transitions are grouped by the qubit at which transition occurs. When transitions are local, each transition is indistinguishable from the other in the same group. A similar diagram can be obtained if $\v{1}\lr\v{e_0}$ transitions are of interest (as in Sec.~\ref{sec:CCZ-tr}) by inverting indexes 0 and 1.
}\end{center}\end{figure}

\begin{figure}\begin{center}
\includegraphics[width=0.6\columnwidth]{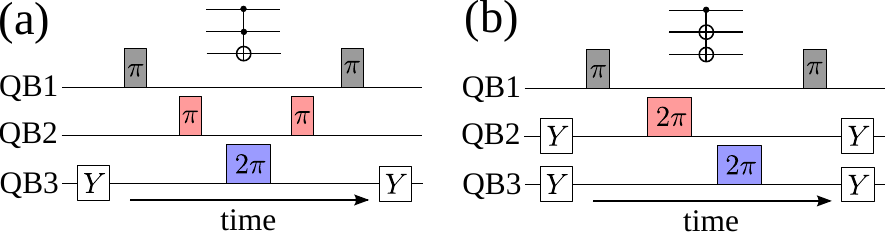}
\caption{\label{fig:3QB-MQB-B}
Three-qubit entangling gates performed via a sequence of pulses. (a) A three qubit Toffoli gate. (b) Cascaded CNOT gate performed by local resonant pulses.
}\end{center}\end{figure}

In this section we demonstrate how entanglement can be manipulated in systems under IRR conditions using a pulse-sequence approach. 

\subsection{Three-qubit CCZ and CZZ gates}

We begin with the simplest multiqubit entangling gate---a three-qubit Toffoli gate \cite{nielsenchuang,Barenco}. The Toffoli gate can be decomposed into
\begin{eqnarray}\label{eq:IRR:Toffoli}
{\rm Toffoli} = H_3 {\rm CCZ}\, H_3,
\end{eqnarray}
or represented by six CNOT gates \cite{Shende}. Here $H_i$ is a Hadamard operation \cite{nielsenchuang} applied to $i$-th qubit, and 
\begin{eqnarray}\label{eq:IRR:CNOT}
{\rm CNOT} = H_2 {\rm CZ}\, H_2
\end{eqnarray}
is a Control-NOT operation. The control-Z gate with one (two) control qubits and one target qubit is denoted by CZ (CCZ). Note that control-Z operations with arbitrary number of control qubits, [C]$^{(n)}$Z, are diagonal identity matrices with, e.g., one diagonal entry set to $-1$. Therefore, to perform the entangling part of the Toffoli gate one must add a phase of $\pi$ to the amplitude residing on, e.g., one of the basis states. The three-qubit basis states are shown in Fig.~\ref{fig:3QB-MQB-A} schematically. All $\v{0}\leftrightarrow\v{e_0}$  transitions in Fig.~\ref{fig:3QB-MQB-A} are grouped by the qubit in which transition takes place. In the IRR all transitions within each group are indistinguishable ($\Delta\omega\to 0$). If we apply a {\it resonant} pulse to transition $\v{000}\lr\v{e_000}$ it will also affect states $\v{001}$, $\v{010}$, and $\v{011}$ in exactly the same fashion (solid lines in Fig.~\ref{fig:3QB-MQB-A}), and, hence, will be completely local to QS-1. However, if a local pulse is applied to, e.g., the first qubit such that some part of the population remains in the excited auxiliary state after the pulse, the next pulse applied to, e.g., the second qubit is no longer local because two of its transitions (from $\v{000}$ and $\v{100}$) affect only a portion of the population. The population left on the excited auxiliary states by the first pulse can not be transfered to higher energy states by the second pulse, because $\Delta\omega'$ is not negligible due to condition (\ref{eq:IRR}).

A three-qubit CCZ gate can be constructed based on three pulses described above. Each pulse induces a transition in a pair of states according to Eq.~(\ref{eq:Lam:2pi}). We choose to apply $\theta=\pi$ and $\theta=2\pi$ pulses as shown in Figs.~\ref{fig:3QB-MQB-B}(a) and (b). Because the combination of two $\pi$ pulses completely returns the population back to the initial state adding a phase of $\pi$, a set of pulses shown in Fig.~\ref{fig:3QB-MQB-B}(a) will change the sign of the amplitudes that reside on all states but $\v{111}$, realizing a CCZ gate (up to a negligible common phase factor). Similarly, pulse sequence shown in Fig.~\ref{fig:3QB-MQB-B}(b) will realize a CZZ gate (two cascaded CZ gates).

\subsection{Multi-qubit entangling gates by a pulse train}

\begin{figure}\begin{center}
\includegraphics[width=0.4\columnwidth]{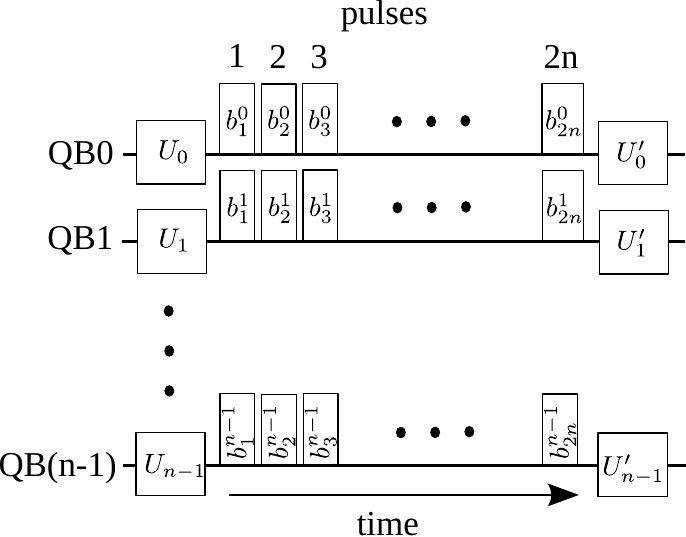}
\caption{\label{fig:3QB-MQB-C}
Pulse scheme for multiqubit entangling gates performed via a sequence of pulses.
}\end{center}\end{figure}

The above pulse-sequence approach can now be easily generalized to larger number of qubits, Fig.~\ref{fig:3QB-MQB-C}, and arbitrary phase accumulation for the states. The latter can be achieved if in any sequence of two consecutive $\pi$ pulses, the pulse frequency is detuned from resonance by some value $\delta$. In the case of $\cosh$-shaped pulses the resulting phase accumulation, $\phi$, is
\begin{eqnarray}\label{eq:IRR:phi}
\exp i\phi = - z/z^*,
\quad\quad\quad
z = \sigma + i\delta
\end{eqnarray}
where $\sigma$ is the pulse bandwidth \cite{rosenzener}. 

If we denote a sequence of $\pi$ pulses applied to $m$-th QS by bits of a binary number 
\begin{eqnarray}\label{eq:IRR:bits}
b^{(m)} = \{b^m_1,b^m_2,...,b^m_{2n}\},
\end{eqnarray}
such that the value $b^m_i = 1$ corresponds to a $\pi$ pulse and $b^m_i = 0$ to free evolution (no pulses), the pulses will not interfere if
\begin{eqnarray}\label{eq:IRR:bit_and}
b^{(1)} \&\, b^{(2)} \&\, b^{(3)} \&\,... = 0,
\end{eqnarray}
where $\&$ is a bit-wise classical AND operation. This set of pulses can perform any diagonal multiqubit gate. For example, $b^{(m)}= 2^m+2^{2n-m-1}$ will perform multiqubit Toffoli, [C]$^{(n-1)}$Z, gate; $b^{(0)}= 2^{2n-1}+1$, $b^{(m>0)}= 3\times 2^{2m-1}$ will accomplish a cascade of CZ gates, or C[Z]$^{(n-1)}$. A single quantum Fourier transform cascade \cite{nielsenchuang} can be performed with the same sequence of pulses as C[Z]$^{(n-1)}$, detuning each pair of $b^{(m>0)}$ pulses from the exact resonance condition to accumulate a phase $\pi/2^{m}$.

\begin{figure}\begin{center}
\includegraphics[width=0.99\columnwidth]{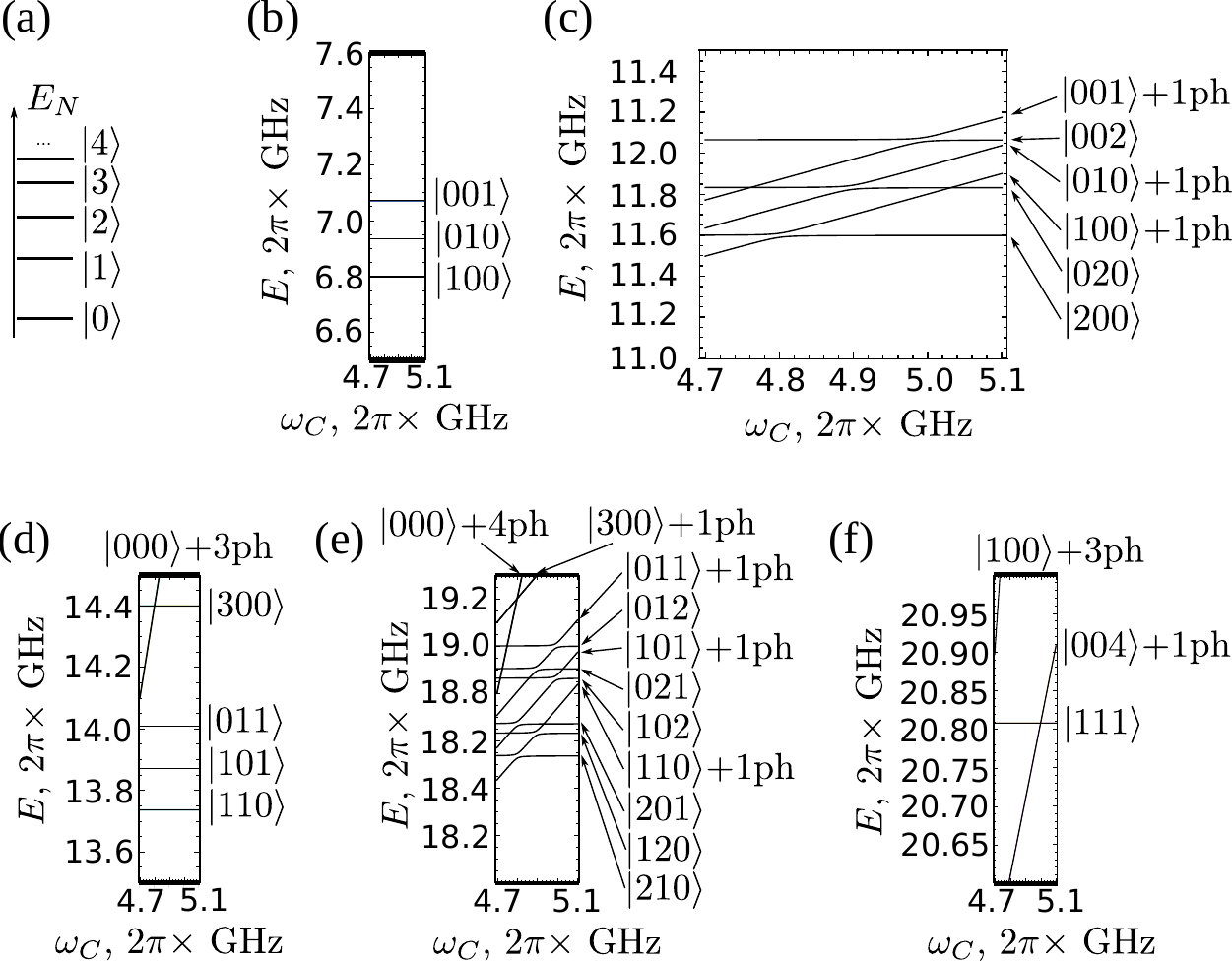}
\caption{\label{fig:3QB-MQB-D}
Numerical simulations of a system of three transmons \cite{Koch-Schoelkopf,SC-review} in a cavity. Schematic energy structure of each transmon is shown in panel (a). Representative parts of the combined numerically-obtained spectrum are shown in panels (b) through (f). The dependence on cavity frequncy $\omega_C$ is emphasized in panel (c) for reference. The transmon frequency is $6.8\times 2\pi$ GHz (anharmonicity is $-2\times 2\pi$ GHz). The frequencies of the second and third transmons are scaled by $1.02$ and $1.04$; $g=10\times 2\pi$ MHz.
% relative anharmonicity alpha_r = alpha/w0 = 0.294 
% the alpha_r is typicaly no larger then 0.1
% which is about 3 times smaller then we have here
}\end{center}\end{figure}

Note that pulse schemes in Figs.~\ref{fig:3QB-MQB-B} and \ref{fig:3QB-MQB-C} are optimized versions of multiqubit gates represented via two-qubit entangling CZ gates, introduced in \cite{Solenov-2QB} for similar architectures. Thus, temporal complexity of the gates still scales as $\O(n)$, that is, approximately $n$ two-qubit CZ gates (or pulses realizing those gates) are needed to perform the operation. Thus, the (temporal) complexity class is the same as in traditional CNOT representation: the best representation of [C]$^{(n-1)}$Z gate via two-qubit entangling gates requires $2n$ individual CZ gates \cite{Shende}. Therefore these optimized pulse sequences can be used to benchmark quantum-walk-based gates introduced further against known CNOT-base decompositions.

\subsubsection{Example: CCZ gate in transmon-cavity system}
\label{sec:CCZ-tr}

As an example of application of the above scheme, we numerically model a system of three transmons \cite{Koch-Schoelkopf,SC-review} interacting via a single microwave cavity mode. 
We model the $b^{(m)}= 2^m+2^{5-m}$ sequence and compute fidelity \cite{comment-Fidel} of the resulting CCZ gate operation. For this example we can describe transmons approximately as anharmonic oscillators [see Fig.~\ref{fig:3QB-MQB-D}(a)] with energies \cite{Koch-Schoelkopf}
\begin{eqnarray}\label{eq:transmon}
E_N = \left(\omega_0 - \frac{\alpha}{2}\right) N
+ \frac{\alpha}{2} N^2,
\end{eqnarray}
and anharmonicity $\alpha$ defined as
\begin{eqnarray}\label{eq:anharmonicity}
\alpha = \omega_1 - \omega_0 < 0,
\end{eqnarray}
where
\begin{eqnarray}\label{eq:omegas}
\omega_0 = E_1-E_0,
\quad\quad\quad
\omega_1 = E_2-E_1.
\end{eqnarray}
The states with $N=0,1,2,3,...$ correspond to states $\v{0}$, $\v{1}$, $\v{e_0}$, $\v{e_1}, ...$, respectively. In transmon qubits the cavity mode (due to either 3D or planar transmission-line cavity) can couple strongly to consecutive transition in each transmon. We chose the frequency of the cavity such, that it is near resonance, but somewhat detuned from $\v{1}\lr\v{e_0}$ transition in each transmon \cite{Koch-Schoelkopf} to realize IRR condition (\ref{eq:IRR}).

As it is the case for all transmon qubit systems, multi-qubit computational basis states are scattered over the large part of the spectrum with multiple other states laying in between due to very small anharmonicity in these systems. In order to correctly account for all possible relevant transitions available in this system we used 5 states to represent each of the three qubits and at least 5 states to represent the cavity mode. Some parts of the complete numerical spectrum (over 600 states) are shown in Figs.~\ref{fig:3QB-MQB-D}(b)-(f), where we plot energies as a function of cavity mode frequency $\omega_C$. States beyond $\v{e_0}$ were not activated by external pulses, but were included to numerically confirm that the population remains at lower energy states when the gate is effective.

Quantum evolution in systems with many accessible transitions across different frequency range becomes less stable numerically. Implicit integration techniques can partially mitigate this problem, however development of appropriate probability-conserving numerical approaches are needed to correctly account for large number of accessible transitions without artificial cutoffs. In order to guarantee acceptable levels of numerical error, we chose anharmonicity that is about three times larger than what is expected for a transmon \cite{Koch-Schoelkopf}. Transmons with large coherence times \cite{Paik} are designed with even smaller anharmonicities, $\alpha\sim 0.2\times 2\pi$ GHz, to suppress charge fluctuation noise \cite{Koch-Schoelkopf}. Due to $2\pi$-periodic conditions on the phase of the superconducting order parameter, the system can, in principle, tunnel to the other identical minima. This tunneling is negligible in the transmon limit\cite{Koch-Schoelkopf} due to large potential barrier separating the minima and the approximate description (\ref{eq:transmon}) is well justified. We adopt this approximation in our numerical modeling to investigate the regime consistent with the experimentally used set of parameters.

\begin{figure}\begin{center}
\includegraphics[width=0.5\columnwidth]{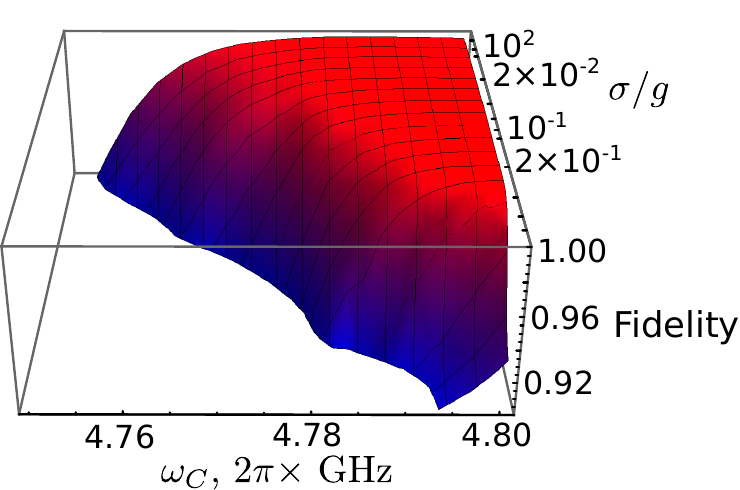}
\caption{\label{fig:3QB-MQB-E}
Numerically computed fidelity of the three-qubit entangling gate CCZ performed using the pulse-train approach outlined in Fig.~\ref{fig:3QB-MQB-C} in a system of three transmons in a cavity (see Fig.~\ref{fig:3QB-MQB-D}). Fidelity is shown as a function of the cavity mode frequency $\omega_C$ and the pulse bandwidth $\sigma$. Parameters are the same as in Fig.~\ref{fig:3QB-MQB-D}.
}\end{center}\end{figure}

In Figure~\ref{fig:3QB-MQB-E} we show numerically computed fidelity of the CCZ gate operation as a function of the cavity mode frequency and the bandwidth of the pulses. A train of four $\pi$ pulses of the same bandwidth $\sigma$, as prescribed by the scheme outlined in Fig.~\ref{fig:3QB-MQB-C}, were taken to realize the gate suggested in Fig.~\ref{fig:3QB-MQB-B}(a). Gaussian pulse shapes with appropriate temporal cut-offs were taken for all pulses. A plateau of high fidelity of the gate operation shrinks and disappears when the pulse bandwidth approaches $g$, and cavity-induced splittings are no longer distinguished by the pulses. This behavior is consistent with the qualitative analysis drawn from the level diagram shown in Fig.~\ref{fig:3QB-MQB-A} and discussed above. At smaller $\sigma$ (not shown) the plateau must descent and eventually vanish when $\sigma$ approaches the decoherence rates.

\section{Overlapping pulses and quantum walks}

Note that in the above examples a sequence of non-overlapping pulses was used. This is an artificial constraint. When a larger Hilbert space is involved and pulses are allowed to overlap, the entire gate can be executed with a single multicolor pulse of comparable strength per Fourier harmonic. This suggests a potential change in the (temporal) complexity class for the multiqubit gates. In what follows, we demonstrate how evolution governed by fully overlapped pulses can be designed to perform entangling quantum gates. 

A network of transitions in a multi-state  multiqubit system can be viewed as a graph, with each node describing a single state, $\v{\psi_n}$, of the diagonalized Hamiltonian (\ref{eq:H_diag}), and edges describing accessible transitions. This graph can have disconnected parts as, e.g., in the case of a three-qubit system with only $\v{i}\leftrightarrow\v{e_i}$ transitions allowed [Fig.~\ref{fig:graphs}(a)]. Quantum evolution through such graph becomes non-trivial when external (control) pulse $V(t)$ has harmonics that enable certain transitions    
\begin{eqnarray}\label{eq:V}
V(t) \sim \Phi(t) \Omega_{000,e_000} \v{000}\iv{e_000} \cos(\omega_{000,e_000} t)
+...
\end{eqnarray}
For simplicity we consider resonant $\omega$ frequencies only. In this case the common pulse profile $\Phi(t)$ enters only via
\begin{eqnarray}\label{eq:tau}
\int_{0}^{t}dt\Phi(t)/\hbar
\to \int_{-\infty}^{\infty}dt\Phi(t)/\hbar
\equiv \tau,
\end{eqnarray}
and the entire evolution operator in the rotating frame is 
\begin{eqnarray}\label{eq:U}
\hat U(t) = \exp[-i\hat\Omega\tau].
\end{eqnarray}
Here $\hat\Omega$ is the complex hermitian adjacency matrix composed of Fourier amplitudes $\Omega$ entering Eq.~(\ref{eq:V}). 

Evolution operator (\ref{eq:U}) describes propagation of amplitude due to a continuous time quantum walk \cite{Farhi,Childs-Spielman,Shenvi,Kempe-REV}---a continuous time quantum evolution in a single-particle quantum system initially set up to occupy only one of the quantum states available. A graph corresponding to such quantum walk is a set of nodes (states) with edges (connections) representing elements in adjacency matrix $\hat\Omega$, which have the physical meaning of dimensionless Rabi frequencies set up by external control pulses. To obtain the desired $\hat U$, corresponding to a quantum gate, e.g., CCZ, in the qubits' computational basis, we analytically solve quantum walk through the graph and chose appropriate $\hat\Omega$ based on that solution.

In order to produce a gate, we need quantum evolution that satisfies specific initial and final conditions: the walk must start and end (at time $\tau$) on one of the qubit basis states. The {\it free parameters} to be adjusted are the amplitudes of the Fourier harmonics of the control pulse (elements of $\hat\Omega$ matrix). The gate is non-trivial if non-trivial phases are accumulated during the walk. Note that Fourier harmonics $\hat\Omega$ play the role of pulse amplitudes if we compare quantum walk approach to the pulse train approach outlined in Fig.~\ref{fig:3QB-MQB-C}. Therefore magnitudes of entries in $\hat\Omega$ must be compared to the magnitudes of pulses in Figs.~\ref{fig:3QB-MQB-B} and \ref{fig:3QB-MQB-C}. The substaintial difference however stems from the fact that when the gate is performed via quantum walks, the ``sequence'' of pulses is ``applied'' in {\it frequency} space rather than in {\it time} and the entire gate is executed in a {\it single} pulse (concurrently). In this case the system is left to explore multiple quantum mechanical trajectories involving entangled states at once, as opposed to being guided (classically) through a single trajectory as it happens for pulse-sequence schemes in Figs.~\ref{fig:3QB-MQB-B} and \ref{fig:3QB-MQB-C}.

In most cases, $\hat\Omega$ and the corresponding matrix, $\hat\omega$, of frequencies $\omega$ entering Eq.~(\ref{eq:V}) are not unique. In what follows we give several specific configurations to illustrate a three-qubit CCZ gate and discuss systems with larger number of qubits in the last section. For the purpose of illustration we discuss a system with only one transition, $\v{0}\lr\v{e_0}$, per qubit system accessible to external control pulses and show $\v{1}\lr\v{e_1}$ transitions [dashed lines in Fig.~\ref{fig:graphs}(a)] to demonstrate the more general structure of the graph. In order to understand how many different Rabi frequencies are available for adjustments, we recall that in systems under the IRR condition (\ref{eq:IRR}), transitions between different levels [$i$,$ii$, and $iii$ in Fig.~\ref{fig:graphs}(a)] are spectrally distinct and can be addressed independently \cite{Solenov-2QB}. Connections within each level correspond to transition in the same QS and are indistinguishable and can not be adjusted independently. The numbering convention for Rabi frequencies in Fig.~\ref{fig:graphs} is set to indicate distinguishable frequencies. 

\subsection{Example: CCZ gate}

\begin{figure}\begin{center}
\includegraphics[width=0.7\columnwidth]{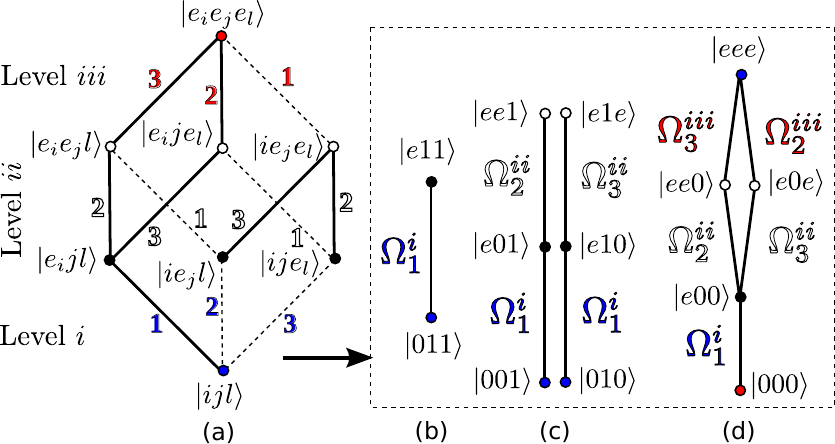}
\caption{\label{fig:graphs}
Networks of states. (a) A complete set of graphs for a three-qubit system. (b-d) Sub-sections of (a) activated to perform CCZ gate. Labels refer to the corresponding Rabi frequencies in $\hat\Omega$. In panels (b), (c), and (d), we drop the subscript in $\v{e_0}$ for clarity. 
}
\end{center}\end{figure}

As an example, we construct a control pulse of five harmonics activating connection 1 on level $i$, and connections 2 and 3 on level $ii$ and $iii$ [solid lines in Fig.~\ref{fig:graphs}(a)]---all are transitions of type $\v{0}\lr\v{e_0}$ in different qubit systems if labeled in the limit $g\to 0$. This produces four non-trivial graphs, each connected to one of the qubit basis states [Fig.~\ref{fig:graphs}(b)-(d)]. A CCZ gate is realized if, e.g., one of the qubit basis states accumulates a phase of $\pi$ for the amplitude residing on it. If we chose $\Omega^{ii}_2 = \Omega^{ii}_3$, the phases accumulated for amplitudes residing on $\v{011}$, $\v{001}$, and $\v{010}$ at time $\tau$ can be easily found from diagonalization of respective adjacency matrices. They are $\Omega^{i}_1\tau = \pi$ for (b), and
\begin{eqnarray}\nonumber
\tau\sqrt{(\Omega^{ii}_{2,3})^2+(\Omega^{i}_1)^2} = 2\pi
\end{eqnarray}
for (c) graphs. All trivial (one state) graphs accumulate no phase (in rotating frame). Two Rabi frequencies, $\Omega^{iii}_2$ and $\Omega^{iii}_3$, must be adjusted to ensure that the phase accumulated for state $\v{000}$ is an even multiple of $\pi$. We can set 
\begin{eqnarray}\nonumber
X = \Omega^{iii}_3/\Omega^{iii}_2 \neq 1,
\end{eqnarray}
transforming the balloon graph (d) into a linear chain of 5 states
\begin{eqnarray}\nonumber
\v{000},
\v{e_000},
(\v{e_0e_00}+\v{e_00e_0})/\sqrt{2},
\v{e_0e_0e_0},
(\v{e_0e_00}-\v{e_00e_0})/\sqrt{2}
\end{eqnarray}
with Rabbi frequencies
\begin{eqnarray}\nonumber
\omega_\alpha &=& \Omega^{i}_1 = \pi/\tau,
\\\nonumber
\omega_\beta &=& \sqrt{2}\Omega^{ii}_2 = \sqrt{6}\pi/\tau,
\\\nonumber
\omega_\delta &=& \Omega^{iii}_2 (1+X)/\sqrt{2},
\\\nonumber
\omega_\gamma &=& \Omega^{iii}_2 (1-X)/\sqrt{2}.
\end{eqnarray}
The corresponding adjacency matrix has
four non-zero eigenfrequencies
\begin{eqnarray}\nonumber
\pm\Omega_\pm = \pm\sqrt{R^2\pm\chi}/\sqrt{2},
\end{eqnarray}
where
\begin{eqnarray}\nonumber
R^2 &=& |\omega_\alpha|^2 + |\omega_\beta|^2 + |\omega_\delta|^2 + |\omega_\gamma|^2
\\\nonumber
(R^4-\chi^2)/4 &=& |\omega_\alpha|^2|\omega_\delta|^2 + |\omega_\beta|^2|\omega_\gamma|^2 + |\omega_\alpha|^2|\omega_\gamma|^2
\end{eqnarray}
with $\Omega_+\neq\Omega_-$. By setting $\tau\Omega_+=2\pi n$ and $\tau\Omega_-=2\pi m$ with integers $n\neq m$, we set the evolution operator $\hat U(\tau)$ for graph (d) to identity. For $n=1$ and $m=2$ we find
\begin{eqnarray}\nonumber
\Omega^{iii}_{2,3}\!\!=\!(3\pm\sqrt{17})\pi/2\tau.
\end{eqnarray}
Note that an arbitrary number of $2\pi$ phase can be added for each $\Omega_\pm\tau$, i.e., $n$ and $m$ can be arbitrary non-equal integers. This means that the choice of Rabi frequencies coming due to external pulse control and realizing CCZ gate via quantum walks is not unique.

\section{Larger number of qubits}

Both, the pulse train and the quantum walks approach, as described above, rely on spectrally distinct states and can suffer from spectral degeneracies and reduction of the effective interaction strength that can occur in larger systems of qubits interacting via a single quantum resonator. Considering a system of transmons as an example, it is natural to assume that the condition that allows to avoid interference due to spectral crowding is
\begin{eqnarray}\label{eq:multi:cond}
|\alpha^{(ij)}| \gtrsim |g|,
\quad\quad\quad
\alpha^{(ij)} = \omega_1^{(i)} - \omega_0^{(j)}
\end{eqnarray}
where superscripts refer to different qubit systems and the frequencies are defined in Eq.~(\ref{eq:omegas}). Note that all pulses are designed to distinguish energies $\sim |g|$ in order to manipulate entanglement. For the purpose of estimating the above condition, we can assume that QS parameters are scaled as $\omega_n^{(i)} = \omega_n (1+ki)$. This leads to $\alpha^{(ij)} = \omega_1 k (i-j) + \alpha (1+kj)$, where we recall that $\alpha<0$.  As the result, condition (\ref{eq:multi:cond}) becomes $||\alpha/\omega_1|-kM| \gtrsim |g/\omega_1|$, where $M$ is the number of qubits. In realistic systems of transmons one can have $|g/\omega_1|\sim 0.001$, $|\alpha/\omega_1|\sim 0.1-0.01$, and can set the scaling factor to about $k\sim 0.01-0.001$. This produces the limit of $M<10$. In addition, the effective interaction must be kept larger than all decoherence rates
\begin{eqnarray}\label{eq:multi:geff}
g_{eff} \sim g \frac{g}{{\frak O}} \gg {\rm max}[\Gamma].
\end{eqnarray}
This limits usable coupling strength between QSs at the opposite ends of the spectrum as $|g/\omega_1|\gg kM\,{\rm max}[|\Gamma/g|]$.

The above limitation on $M$ stems from the fact that we have considered a simplified model to introduce a new approach for entanglement-manipulating gates based on quantum walks. Specifically, the operation of the gate was demonstrated analytically based on spectral selectivity that relies on frequency detuning. In actual physical systems selectivity of pulses is determined not only by detuning, but also by the magnitude of the transition matrix elements present for the given pulse field distribution and the wave functions involved in the transition of interest. In other words, if the pulse field does not overlap with the states involved in the given transition it does not affect that transition even if it is in or near resonance with the pulse frequency. This makes transitions corresponding to different edges of the graph, e.g., in Fig.~\ref{fig:graphs}(a), individually addressable even when the corresponding transition frequencies are near resonance with each other. Furthermore, in many architectures scalable qubit registers would involve multiple cavities carrying interaction between different parts of the system, rather then a single quantum resonator coupled to every QS. This introduces larger number of addressable auxiliary states, hence generating larger graphs with addressable edges. As the result, restrictions (\ref{eq:multi:cond}) and (\ref{eq:multi:geff}) no longer apply to the entire qubit register, although, they can still be important locally. This situation is ideal for the proposed approach of using continuous time quantum walks to manipulate entanglement in multiqubit systems. Investigation of quantum walk properties in this context, particularly ``return'' quantum walks discussed here, is an interesting direction of further research, which, however, goes beyond the scope of the present paper. Finally, we note that the evolution through networks of auxiliary states is also affected by decoherence. This can potentially limit sizes of the graphs involved in entangling gates. At the same time, small decoherence  can, in some cases, enhance quantum propagation through networks \cite{Kendon-Tregenna,Solenov-R1,Solenov-R2,Solenov-R3}.

\section{Acknowledgments}

I am grateful to Q. Khan, W. D. Thacker, and D. S. Wisbey for helpful discussions. This work was supported by SLU and SLU HPC.

\end{document}